\documentclass[aps,preprint,prd,epsfig,axodraw,nofootinbib]{revtex4}
\usepackage{amssymb}
\usepackage{amsmath}
\usepackage{graphicx}
\usepackage{fancyhdr}
\usepackage{epsfig}
\newcommand{\be}{\begin{equation}}
\newcommand{\ee}{\end{equation}}
\newcommand{\bea}{\begin{eqnarray}}
\newcommand{\eea}{\end{eqnarray}}
\newcommand{\bes}{\begin{subequations}}
\newcommand{\ees}{\end{subequations}}

\newcommand{\bc}{\begin{center}}
\newcommand{\ec}{\end{center}}

\begin{document}

\title{Explaining ATLAS and CMS Results Within the Reduced Minimal 3-3-1 model}

\author{W. Caetano}
\affiliation{{ Departamento de
F\'{\i}sica, Universidade Federal da Para\'\i ba, Caixa Postal 5008, 58051-970,
Jo\~ao Pessoa, PB, Brasil}}

\author{D. Cogollo}
\affiliation{{Departamento de
F\'{\i}sica, Universidade Federal de Campina Grande, Caixa Postal 10071, 58109-970,
Campina Grande, PB, Brasil }}
\author{C. A. de S. Pires}
\affiliation{{ Departamento de
F\'{\i}sica, Universidade Federal da Para\'\i ba, Caixa Postal 5008, 58051-970,
Jo\~ao Pessoa, PB, Brasil}}

\author{Farinaldo  S. Queiroz}
\affiliation{{ Santa Cruz Institute for Particle Physics, University of California, 1156 High St., Santa Cruz, CA 95064, USA}}

\author{P. S. Rodrigues da Silva}
\affiliation{{ Departamento de
F\'{\i}sica, Universidade Federal da Para\'\i ba, Caixa Postal 5008, 58051-970,
Jo\~ao Pessoa, PB, Brasil}}

\date{\today}

\begin{abstract}
Recently the ATLAS and CMS collaborations announced the discovery of a higgs particle with a mass of $\sim 125$~GeV. The results are mildly consistent with the Standard Model Higgs boson. However, the combined data from these collaborations seem to point to an excess in the $h \rightarrow \gamma \gamma$ channel. In this work we analyze under which conditions this excess may be plausibly explained within the reduced minimal 3-3-1 model, while being consistent with $b\overline{b}$, $WW$, $ZZ$ and $\tau^{+} \tau^{-}$ channels. Moreover, we derive the properties of the heavy neutral and the doubly charged scalars predicted by the model. We then conclude that at a scale of a few TeV, this model provides a good fit to the ATLAS and  CMS  signal strength measurements, and therefore stands as an appealing alternative to the standard model.
\\
%
\end{abstract}

\maketitle

\section{Introduction}

Higgs Hunters around the world have been excited since the resonance found at $\sim 125$ GeV reported by the Atlas \cite{atlas} and CMS \cite{cms} collaborations at 5~$\sigma$, in July 2012. The results obtained in different search channels were mildly consistent with a standard model (SM) Higg-like boson. Interestingly, both collaborations reported an excess of events in the diphoton channel ($h \rightarrow \gamma \gamma$), which under the no background hypothesis, could be explained in beyond SM models only~\cite{explaining}. However, those collaborations have been gathering data since then, and recently they have updated their former results with better statistics at the Moriond Conference \cite{RencontresdeMoriond}. 

It is opportune to remind  the existence of a simple gauge extension of the SM, namely  $SU(3)_c\otimes SU(3)_L\otimes U(1)_N$~\cite{331-1,331-2}, that is a very competitive alternative to the well tested SM, which  addresses important problems from a  theoretical perspective such as electric charged quantization \cite{ecq}, the neutrino's masses \cite{neutrinos1}, family puzzle \cite{families},  as well as from a phenomenological point of view, such as dark matter \cite{DM331}, Higgs search \cite{higgs331}, mesons oscillations \cite{novelsourses} and  exotic particles \cite{exotic}. There are many variations of this electroweak extension proposed in the literature and they have their own phenomenology and interesting aspects \cite{dong}. Concerning the minimal version of the model, the so called minimal 3-3-1 model, where all known leptons are in a fundamental representation of the $SU(3)_L$ group, it was shown that its scalar content, originally three triplets and one sextet, can be reduced to only two scalar triplets~\cite{paulodias}. This reduced minimal 3-3-1 (RM331) model is relatively interesting for having a short scalar sector when compared with other non-abelian gauge extensions of the SM\cite{pheno}. After spontaneous symmetry breaking to $SU(3)_C \otimes U(1)_{QED}$, the physical scalar spectrum of the RM331 model is composed solely by three scalars, where two of them are neutral CP-even scalars with the third one being electrically  doubly charged. One of the neutral scalars ought to play the role of the SM Higgs-like boson. 
We then identify our Higgs-like boson and, by deriving all relevant interactions involving it in order to check under which conditions it fits the experimental results, we observe that it has similar properties to the SM Higgs boson.  To do so,  we compare our results with the most recent CMS and ATLAS analyses regarding the signal strength with focus on the $h \rightarrow \gamma \gamma$ channel.  Additionally, for completeness, we will obtain the interactions and branching ratios of the remaining neutral and the doubly charged scalars. 

This work is divided as follows: In Sec.~(\ref{themodel}), we will present the RM331 model summarizing its key features, including the mass spectrum. In Sec.~(\ref{Higgsdecay}) we calculated the scalars' decay channels, including the heavy ones and the doubly charged.  In Sec.~(\ref{analysis}), we will perform a numerical analyses of the Higgs-like particle with focus on the $\overline{l}l$, $\overline{q}q$, $WW^{*}$,$ZZ^{*}$, and $\gamma \gamma,$ channels. Lastly, in Sec.~(\ref{conclusions}) we draw our conclusions.

\section{The RM331 Model}
\label{themodel}
The leptonic content of the RM331 model  is composed by the following triplet,
\begin{equation}
\begin{array}{cc}
f_L=\left(\begin{array}{c}
\nu_l \\  l\\ l^c
\end{array}\right)_L \sim (1,3,0),
\end{array}
\label{leptonsector}
\end{equation}
where $ l=e,\mu ,\tau.$ The numbers between parentheses refer to the $SU(3)_C$, $SU(3)_L$, $U(1)_N$ quantum numbers, respectively.\\

In the quark sector, anomalies cancellation require that  one generation comes in a $SU(3)_L$ triplet and the other two come in anti-triplet representation as follows,
\begin{eqnarray}\label{quarkcontent}
&&Q_{1L}  =  \left(\begin{array}{c}
u_1\\
d_1\\
J_1
\end{array}\right)_L \sim (3,3,+\frac{2}{3}),\,\,\,
Q_{iL}  =  \left(\begin{array}{c}
d_i\\
-u_i\\
J_i
\end{array}\right)_L \sim (3,3^*,-\frac{1}{3}), \nonumber \\
&&\begin{array}{ccc}
u_{iR} \sim (3,1,+\frac{2}{3}); & d_{iR}
\sim(3,1,-\frac{1}{3}); & J_{iR}
\sim(3,1,-\frac{4}{3}),
\end{array} \nonumber \\
&&\begin{array}{ccc}
u_{1R} \sim (3,1,+\frac{2}{3}); & d_{1R}
\sim(3,1,-\frac{1}{3}); & J_{1R}
\sim(3,1,+\frac{5}{3}),
\end{array}
\end{eqnarray}
with $i=2,3$.

The scalar sector is composed by two scalar triplets only, 
\begin{equation}\label{fields}
\rho = \left(\begin{array}{c} 
\rho^{+}\\
\rho^{0}\\
\rho^{++}
\end{array}\right) \sim (1,3,1), \quad
\chi = \left(\begin{array}{c} 
\chi^{-}\\
\chi^{--}\\
\chi^{0}
\end{array}\right) \sim (1,3,-1).
\end{equation}

It has been shown that these two triplets are sufficient to break the $SU(3)_C \times SU(3)_L \times U(1)_N$ symmetry to the $SU(3)_C \times U(1)_{QED}$, and also generate the correct masses of all fermions and gauge bosons~\cite{paulodias}.

These scalars allow us to write down the most general gauge and Lorentz invariant potential as,
\begin{eqnarray}\label{potential}
V(\chi, \rho) &=& \mu_1^2 \rho^{\dagger}\rho + \mu_2^2 \chi^{\dagger}\chi + \lambda_1(\rho^{\dagger}\rho)^2 + \lambda_2 (\chi^{\dagger}\chi)^2 \nonumber\\
&&+ \lambda_3 (\rho^{\dagger}\rho)(\chi^{\dagger}\chi) + \lambda_4(\rho^{\dagger}\chi)(\chi^{\dagger}\rho).
\end{eqnarray}

One might have noticed that we have two neutral scalars in Eq.~(\ref{fields}), which can develop a nontrivial vacuum expectation value (VEV). Therefore we shift these fields accordingly, 
\begin{equation}
\rho^0, \chi^0 \rightarrow \dfrac{1}{\sqrt{2}}(v_{\rho,\chi} + R_{\rho,\chi} + iI_{\rho,\chi}),
\end{equation}
where  by supposing that $v_{\chi} \gg v_{\rho}$, we get the following  pattern of spontaneous symmetry breaking (SSB), 
$$ 
{\rm SU(3)}_L \otimes {\rm
U(1)}_X \stackrel{\langle\chi^0\rangle} \longrightarrow {\rm
SU(2)}_L\otimes{\rm U(1)}_Y \stackrel{\langle \rho^0\rangle}
\longrightarrow {\rm U(1)}_{\rm QED}.
$$ 

The set of constraint equations derived from the potential above are,
\begin{eqnarray}
\mu_{1}^2 + \lambda_1 v_{\rho}^2 + \dfrac{\lambda_3}{2}v_{\chi}^2 &=&0, \nonumber\\ 
\mu_{2}^2 + \lambda_2 v_{\chi}^2 + \dfrac{\lambda_4}{2}v_{\rho}^2 &=&0.
\end{eqnarray}
With these constraints,  we obtain the following mass matrix for the CP-even neutral scalars in the basis $(R_\chi\,,\,R_\rho)$,
\begin{equation}
m^2_R=\frac{v^2_\chi}{2}
\begin{pmatrix}
2\lambda_2\ & \lambda_3 t  \\
\lambda_ 3 t& 2\lambda_1 t^2 
\end{pmatrix}
\label{neutralmassmatrix},
\end{equation}
where $t=\frac{v_\rho}{v_\chi}$.  Diagonalizing this mass matrix, we obtain the following eigenvalues,
\begin{equation}
m^2_{h_1} = \left(\lambda_1 - \dfrac{\lambda_3^2}{4\lambda_2}\right)v_\rho^2, \quad
m^2_{h_2} = \lambda_2 v_\chi^2 + \dfrac{\lambda_3^2}{4\lambda_2} v_\rho^2,
\label{massadosneutros}
\end{equation}
and the respective eigenstates, 
\begin{equation}
h_1 = c_\beta R_{\rho} - s_\beta R_\chi, \quad h_2 = c_\beta R_{\chi} + s_\beta R_\rho,
\label{neutralstates}
\end{equation}
where,
\begin{equation}
c_\beta \equiv \cos(\beta) \approx 1 - \dfrac{\lambda_3^2}{8\lambda_2^2}\dfrac{v_\rho^2}{v_\chi^2} \quad \mbox{and} \quad s_\beta \equiv \sin(\beta) \approx \dfrac{\lambda_3}{2\lambda_2}\dfrac{v_\rho}{v_\chi}.
\label{betaangle}
\end{equation} 

It is important to emphasize that for $v_\chi \gg v_\rho$,  which is the regime assumed here, we get $m^2_{h_1} < m^2_{h_2}$ and $c_{\beta} > s_{\beta}$. 

In regard to the CP-odd scalars, $I_\rho$ and $I_\chi$, both are Goldstone bosons  that will be eaten by the gauge bosons $Z$ and $Z^{\prime}$, respectively.

As ATLAS and CMS collaborations announced the discovery of a neutral scalar with a mass of $\simeq 125$~GeV, we have a new constraint over the parameters of the scalar potential. Hence, by taking $v_{\rho}=246$~GeV, we identify the  lightest scalar as the Higgs-like boson, then the above expression to $m_{h_1}$ imposes the following constraint over the potential couplings, $\lambda_{1,2,3}$ ,   
\begin{equation}
\lambda_1 - \dfrac{\lambda_3^2}{4\lambda_2} \approx \dfrac{1}{4},
\end{equation}
which will be respected throughout this work.

For the doubly charged scalars in the basis $(\chi^{++}\,,\,\rho^{++})$, we find the following mass matrix,
\begin{equation}
m^2_{++}=\frac{\lambda_4 v^2_\chi}{2}
\begin{pmatrix}
t^2\ & t  \\
t& 1 
\end{pmatrix},
\label{++massmatrix}
\end{equation}
which has the following eigenvalues,
\begin{equation}
m^2_{\tilde h^{++}}=0\,\,\,\,\,\,\,\mbox{and}\,\,\,\,\,\,m^2_{h^{++}}=\frac{\lambda_4}{2}(v^2_\chi + v^2_\rho),
\label{masscharged}
\end{equation}
and the corresponding eigenvectors,
\begin{eqnarray}
&&\left( 
\begin{tabular}{c}
$\tilde h^{++}$ \\ 
$h^{++} $
\end{tabular}
\ \right) = \left( 
\begin{tabular}{cc}
$c_ \alpha$ & -$s_ \alpha$\\ 
$s_\alpha$ & $c_\alpha$
\end{tabular}
\ \right) \left( 
\begin{tabular}{l}
$\chi^{++} $ \\ 
$\rho^{++} $
\end{tabular}
\right).
\label{chargedeigenvectors}
\end{eqnarray}
where,
\begin{equation}
c_\alpha = \frac{v_\chi}{\sqrt{v^2_\chi + v^2_\rho}}\,,\, \quad s_\alpha=\frac{v_\rho}{\sqrt{v^2_\chi + v^2_\rho}}.
\label{chargedeiggenvectors}
\end{equation}

Once $v_\chi \gg v_\rho$, then $\tilde h^{++} \approx \chi^{++}$ and $h^{++}\approx \rho^{++}$, where $\tilde h^{\pm \pm}$ are the goldstones eaten by the gauge bosons $U^{\pm \pm}$, while $h^{\pm \pm}$ remain as physical scalars in the spectrum and will be referred as doubly charged scalars hereafter.

The simply charged scalars $\rho^{+}$ and $\chi^{+}$ are both Goldstone bosons and are eaten by the gauge bosons $W^{+}$ and $V^{+}$, respectively. 

Regarding the gauge bosons, the kinetic terms are,
\begin{equation}
\mathcal{L=}\left( \mathcal{D}_{\mu }\chi \right) ^{\dagger }\left( \mathcal{%
D}^{\mu }\chi \right) +\left( \mathcal{D}_{\mu }\rho \right) ^{\dagger
}\left( \mathcal{D}^{\mu }\rho \right) ,
\label{dc}
\end{equation}
with
\begin{equation}
D_\mu = \partial_\mu -ig W^a_\mu \frac{\lambda ^a}{2}-ig_N N W^N_\mu,
\label{derivativecovariant}
\end{equation}
where  $a=1,...,8$ and $\lambda^a$ are the Gellmann matrices. After SSB, Eq.~(\ref{dc})
provides the following mass terms for the three standard gauge bosons, namely $W^{\pm}$ and $Z$ ,and for the five new gauge bosons,  $V^\pm$, $U^{\pm \pm}$, and  $Z^{\prime}$,
$$ M^2_{W^{\pm}} = \dfrac{g^2 v^2_{\rho}}{4}, \quad M^2_{Z} = \dfrac{g^2 v^2_{\rho}}{4c^2_W} , \quad M^2_{V{^\pm}} = \dfrac{g^2 v^2_{\chi}}{4},$$

$$\quad M^2_{Z^{\prime}} =\left( \dfrac{g^2 c^2_W}{3-4 s^2_W}\right) v^2_{\chi} \quad \mbox{and} \quad M^2_{U{^{\pm \pm}}} = \dfrac{g^2 (v^2_{\rho} + v^2_{\chi})}{4}, $$
where $c_W=\cos \theta_W$ and $s_W=\sin \theta_W$ with $\theta_W$ being the Weinberg angle.
The interactions involving the scalars and the gauge bosons derived from the kinetic term above are  presented in the appendix, tables~\ref{tableh1} and ~\ref{tableh2}.

In the RM331 Model, the charged lepton masses arise from the effective dimension five operator,
\begin{equation} \label{effective}
\dfrac{\kappa_l}{\Lambda}(\overline{f^c_L}\rho^*)(\chi^\dagger f_L) + h.c.\,,
\end{equation}
where $l = e, \mu, \tau$, are the lepton family labels, $\kappa_l$ is a dimensionless parameter and $\Lambda$, which lies about $\Lambda=4-5$~TeV, is the highest energy scale available where the model loses its perturbative  behavior \cite{landau}. After SSB, this operator generates the following mass term for the charged leptons, 
$$m_{l} \approx \dfrac{\kappa_l v_{\rho} v_{\chi}}{2 \Lambda}.$$ 

As for the standard quarks, their masses come from a combination of renormalizable Yukawa interactions and dimension-five operators,
\begin{eqnarray}
&&
 \lambda^d_{1a}\bar Q_{1L}\rho d_{aR} + \frac{\lambda^d_{ia}}{\Lambda}\varepsilon_{nmp}\left(\bar Q_{iLn}\rho_m\chi_p\right)d_{aR} + \nonumber \\
 && \lambda^u_{ia}\bar Q_{iL}\rho^* u_{aR} + \frac{\lambda^u_{1a}}{\Lambda}\varepsilon_{nmp}\left(\bar Q_{1Ln}\rho^*_m\chi^*_p\right)u_{aR} + h.c.\,,
 \label{quarksmassterms}
\end{eqnarray}
where, again, the $\lambda$s are dimensionless parameters.
The mass terms and the renormalizable interactions among  flavor quark eigenstates and scalars are displayed in the appendix.


\section{Scalar Decays}
\label{Higgsdecay}

The main goal of this work is to study the total decay width of the scalars of the model and their Branching Ratio (BR) in different channels in the framework of the RM331 model. 
Concerning the Higgs boson, we investigate how our results differ from the SM ones. To do it so, we ought to  obtain the signal strength ($\mu$) for each individual channel of the  Higgs. Although in the RM331 model there are no significant new contributions to the production of our Higgs-like boson ($h_1$), the form of the interactions among  $h_1$ and the SM quarks are a bit different  from  the SM case, due to the mixing between $h_1$ and $h_2$, see Eq.~(\ref{neutralstates}). Consequently, the expressions for the width decay rates  turn out to be different from the SM Higgs boson as well. 
It is opportune to  remind that Higgs decay widths in the context of the minimal 331 model were already computed in Ref.~\cite{higgs331}. Nevertheless, once the latter spectrum is larger than the RM331 model, some simplifying assumptions were made on the scalar mixing matrix that severely restricted the parameter space of that model. Here, no such a constraint exists, since there are only two physical neutral scalars, and their BR may vary according to the choice of the mixing angle $\beta$.
Hence,  the signal strength in the RM331 model should include both, production and decay rates of the Higgs-like boson particle,
\begin{equation} \label{signal}
\mu_{xy} = \dfrac{\sigma_{331}(pp \rightarrow h_1)}{\sigma_{SM}(pp \rightarrow h)} \dfrac{BR_{331}(h_1 \rightarrow xy)}{BR_{SM}(h \rightarrow xy)},
\end{equation}
for any final states $x$ and $y$. Here, we are going to consider gluon fusion as the dominant process in the Higgs production. In view of this, as the coupling $\overline{t}th_1$ in the RM331 model differs from the SM by a $c_{\beta}$, then we are going to have the factor: $\sigma_{331}(pp \rightarrow h_1)/\sigma_{SM}(pp \rightarrow h) \sim c^2_{\beta}$ changing the production.  As usual, we are going to consider that  the total Higgs width decay expression  in the signal strength is composed by the following decays:
\begin{equation} \label{totalhiggs}
\Gamma(h_1\rightarrow all)  =  \Gamma_{(h_1 \rightarrow  \bar{l}l ,\overline{q}q )} + \Gamma_{(h_1 \rightarrow ZZ^{*})}  + \Gamma_{(h_1 \rightarrow WW^{*})} + \Gamma_{(h_1 \rightarrow \gamma \gamma)} + \Gamma_{(h_1 \rightarrow gg)}\,. 
\end{equation}
That being said, we will derive now the signal strength in our model for each relevant final state.

\subsection{Higgs Decay into Fermions}

In the RM331 model the width decay of Higgs-like boson $h_1$ into a pair of leptons is found to be,
\begin{eqnarray}
\Gamma_{331}(h_{1} \rightarrow \overline{\ell}\ell) &&=  \dfrac{g^2}{32 \pi}\dfrac{m_\ell m_{h_{1}}}{m_W^2}\left( 1 - \dfrac{4m_{\ell^2}}{m_{h_{1}^2}} \right)^{3/2} \left(c_\beta - \dfrac{v_\rho}{v_\chi}s_\beta\right)^2.
\label{h01ll}
\end{eqnarray}
   
From Eqs.(\ref{signal})-(\ref{h01ll}), we find that the signal strength for this channel is,
\begin{equation} \label{mull}
\mu_{\ell\ell}= c^2_\beta \left(c_\beta - \frac{v_\rho}{v_\chi}s_\beta\right)^{2} 
\dfrac{\Gamma_{SM}(h \rightarrow all)}{\Gamma_{331}(h_1 \rightarrow all)}.
\end{equation}

Observe that, due to the mixture between $h_1$ and $h_2$ given in Eq. (\ref{neutralstates}), the  Eqs.~(\ref{h01ll})-(\ref{mull}) depend on the mixing angle $\beta$.  Thus, in the limit $v_{\chi}\gg v_{\rho}$ (while the scalar couplings are naturally of order of one), $\beta \rightarrow  0$, and $h_1$  recovers the result of the standard Higgs $h$. 

As the masses of the quarks arise from different sources, see Eq. (\ref{quarksmassterms}), the interactions among $h_1$ and the down quarks get different from the up quarks. Due to this, we are going to have two different expressions for the $h_1$ decay width into quarks, namely, 
\begin{eqnarray}
\Gamma_{331}(h_{1} \rightarrow \overline{q}q) &&=  \dfrac{3 g^2}{32 \pi}\dfrac{m_q m_{h_{1}}}{m_W^2}\left( 1 - \dfrac{4m_{q}}{m_{h_{1}^2}} \right)^{3/2}  c^2_\beta, 
\label{h01qq}
\end{eqnarray}
\begin{eqnarray}
\Gamma_{331}(h_{1} \rightarrow \overline{q^{\prime}}q^{\prime}) &&=  \dfrac{3 g^2}{32 \pi}\dfrac{m_q^{\prime} m_{h_{1}}}{m_W^2}\left( 1 - \dfrac{4m_{q^{\prime}}}{m_{h_{1}^2}} \right)^{3/2} \left(c_\beta + \dfrac{v_\rho}{v_\chi}s_\beta\right)^2,
\label{h01qq1}
\end{eqnarray}
for $q=d,c,t$ and $q^{\prime}=u,s,b$.

Finally, from Eqs.~(\ref{signal})-(\ref{totalhiggs}) and Eqs.~(\ref{h01qq})-(\ref{h01qq1}), we get the following expressions for the  signal strength,
\begin{equation} \label{muqq}
\mu_{qq}= c^4_\beta \dfrac{\Gamma_{SM}(h \rightarrow all)}{\Gamma_{331}(h_1 \rightarrow all)},
\end{equation}
\begin{equation} \label{q1q1}
\mu_{q^{\prime}q^{\prime}}= c^2_\beta \left(c_\beta + \frac{v_\rho}{v_\chi}s_\beta\right)^{2} \dfrac{\Gamma_{SM}(h \rightarrow all)}{\Gamma_{331}(h_1 \rightarrow all)}.
\end{equation} 
\\
Once we have obtained the signal strength of our Higgs-like scalar into fermions, we  should next analyze the case where we have either WW or ZZ as a final state.  

\subsection{Higgs Decay into WW and ZZ}

A Higgs-like particle of $\simeq 125$~GeV can decay into $WW^{\star}$ and $ZZ^{\star}$˜\footnote{Here $W^*$ and $Z^*$ refer to the respective virtual gauge bosons.}, followed by gauge boson decays into a pair of leptons or quarks. Following the procedure in Ref.~\cite{rizzo}, we must sum over all possible final states to find, according to the $h_1$ couplings in table~\ref{tableh1} in the appendix,
\begin{eqnarray}
&&\Gamma_{331}(h_1 \rightarrow Z^{\star}Z) = c^2_{\beta}  \dfrac{g^4m_{h_1}}{2048\pi^2}   \left( \dfrac{7 -\frac{40}{3}s^2_{W} + \frac{160}{9}s^4_{W}}{c^4_{W}}\right) F\left(\frac{m_Z}{m_{h_1}}\right),\nonumber \\
&& \Gamma_{331}(h_1 \rightarrow W^{\star}W) = c^2_{\beta}  \dfrac{3g^4m_{h_1}}{512\pi^2} F\left(\frac{m_W}{m_{h_1}}\right),
\label{eqhvirtual}
\end{eqnarray}
where,
\begin{eqnarray}
F(x) = &-& |1-x^2|\left(\frac{47}{2}x^2 - \frac{13}{2} + \frac{1}{x^2}\right) \nonumber \\
&+& 3(1-6x^2+4x^4)|\ln(x)| \nonumber\\ 
&+& \dfrac{3(1-8x^2+20x^4)}{\sqrt{4x^2-1}}\cos^{-1}\left(\dfrac{3x^2-1}{2x^3}\right).
\label{eqhvirtual3}
\end{eqnarray}

This allows us to obtain the signal strength in the RM331 model for $h_1$ into a pair of massive SM gauge bosons  according to, 
\begin{eqnarray}\label{muH1}
\mu_{WW^{\star},ZZ^{\star}} & = & c^4_\beta \dfrac{\Gamma_{SM}(h \rightarrow all)}{\Gamma_{331}(h_1 \rightarrow all)}\,,
\end{eqnarray}
 which in the limit $\beta \rightarrow 0$, i.e. when $h_1\equiv h$, then we have  $\mu_{WW^{\star}}, \mu_{ZZ^{\star}} \rightarrow 1$.

As a nice result, the RM331 model is able to reproduce the CMS and ATLAS results concerning the WW and ZZ channels while enhancing the $2\gamma$, as we shall see further. 

\subsection{Higgs decay into two photons}

\begin{figure}[!]
\includegraphics[width=0.6\columnwidth]{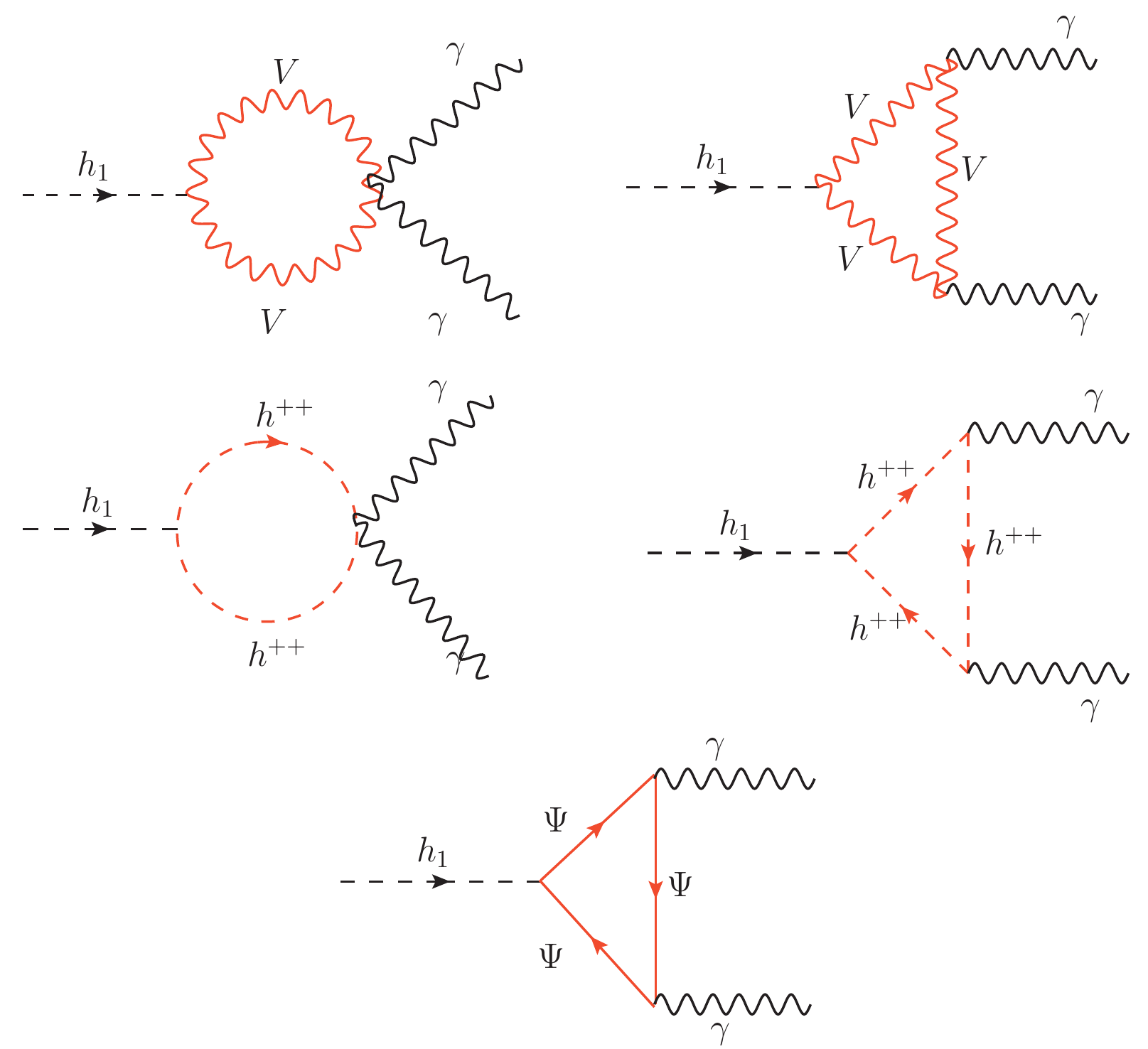}
\caption{The one-loop diagrams contributing to the $h_1 \rightarrow \gamma \gamma$ decay for the charged gauge bosons, $V = U^{++}$, $V^{+}$ and $W^{+}$, the doubly charged scalar, $h^{++}$, and the charged fermions $\Psi$.}
\label{diagram}
\end{figure}

It is well known that the Higgs couples to the photon via loop induced processes. The most relevant diagrams in the SM for $h\rightarrow \gamma \gamma$ are the ones mediated by the top quark, the tau lepton and the gauge boson $W^{\pm}$. Once the RM331 model has an extended scalar and gauge content,  additional contributions to the diphoton decay channel are expected.  These new contributions are due to the extra charged vector bosons $V^{\pm}\,,U^{\pm \pm}$ and the doubly charged scalar $h^{\pm \pm}$, and the main contributions to the decay $h_1 \rightarrow \gamma \gamma$ are shown in Fig.~(\ref{diagram}).
 In order to easily account for new contributions that arise in the decay width into two photons, $\Gamma(h_1 \rightarrow \gamma \gamma)$, we usually write it as~\cite{higgs331,GeneralL},
\begin{equation}
\Gamma_{331}(h_1 \rightarrow \gamma \gamma) = \dfrac{\alpha^2 g^2}{1024\pi^3} \dfrac{m_{h_{1}}^3}{m_W^2} \mid \sum_i  N_c  e_i F_{i} \mid^2\,, 
\label{2gamma}
\end{equation}
where the sum is over all  charged scalars, fermions and bosons of the theory, $e_i$ is the electric charge of the corresponding particle  which runs in the loop, and  $F_{i}$ are form factors  that contain the information about new physics.  These form factors have  analytic expressions for each type of particle running in the loop, and can be written as follows,
\begin{eqnarray}
F_{\phi_i} &=& [\tau_{\phi_i} (1 - \tau_{\phi_i} I^2)]\dfrac{M_{\phi_i}^2}{m_{\phi_i}^2}\ , \nonumber \\
F_{\psi_i} &=& -2 \tau_{\psi_i} [1 + (1 - \tau_{\psi_i}) I^2]\dfrac{M_{\psi_i}}{m_{\psi_i}} \ , \nonumber \\
F_{V_i} &=& [2 + 3 \tau_{V_i} + 3\tau_{V_i} (2 - \tau_{V_i})I^2]\dfrac{m^2_W}{m^2_{V_i}}c_{V_i}, \ 
\label{coefficients}
\end{eqnarray}
where  $m_{\phi_i}$, $m_{\psi_i}$, $m_{V_i}$ correspond to the masses of the charged particles, $c_{V_i}$ are given in the table~\ref{table3cv}.  The mass coefficients $M^2_{\phi_i}$ and $M_{\psi_i}$  are given by,  
\begin{eqnarray}
M^2_{\phi_i} = m^2_{h^{++}} \dfrac{v^2_{\rho}}{(v_{\chi}+v_{\rho})^ 2} (c_\beta - \dfrac{v_\rho}{v_\chi}s_\beta) \,\,,\,\, M_{\psi_i} = m_{(l, q^{\prime})}(c_\beta - \dfrac{v_\rho}{v_\chi}s_\beta),
\label{newmasses}
\end{eqnarray}
where all leptons (as well the quarks $q^{\prime} = u,s,b$) have the same coefficients , while the other quarks $q = d,c,t$ have the following $M_{\psi_i} = m_{q}c_\beta$.

The function $I(\tau_i)$ is given by,
\be
I(\tau_i) \equiv \left\{
\begin{array}{l}
arcsin\left(\sqrt{\frac{1}{\tau_i}}\right)\,\,\,\hbox{for} \,\,\,\tau_i\geq 1 \\
\frac{1}{2}\left[\pi+\imath \ln\left[\frac{1+\sqrt{1-\tau_i}}{1-\sqrt{1-\tau_i}}\right]\right]\,\,\,\hbox{for} \,\,\,\tau_i\leq 1
\end{array}\right.
\ee
where,
$$
\tau_i = \dfrac{4m_i^2}{m^2_{h_1}}\,,
$$
with $m_i$ being the mass of the corresponding particle in the loop.
Notice that the electric charge factor $e_i$ in Eq.~(\ref{2gamma}) may potentially enhance the contributions from  $U^{\pm \pm}$ and  $h^{\pm \pm}$, once they are doubly charged particles,  but such contributions are opposite in sign and some destructive interference may occur. Furthermore, the relevant fermion contribution to the $\Gamma(h_1 \rightarrow \gamma \gamma)$ in the RM331 model comes from the top quark, as no exotic new quark directly couples to $h_1$. 

\begin{table}[h]
\centering
\caption{The $c_{V_i}$ coefficients of third line of Eq.~($\ref{coefficients}$).}
\begin{tabular}{|c|c|c|c|}
\hline 
Higgs & $c_W$ & $c_V$ & $c_U$\\
\hline 
$h_1$ & $c_\beta$ & $-\frac{v_{\chi}}{v_{\rho}}s_\beta$ & $(c_\beta  - \frac{v_\chi}{v_\rho}  s_\beta)$\\
$h_2$ & $s_\beta$ & $\frac{v_{\chi}}{v_{\rho}}c_\beta$ & $(s_\beta  + \frac{v_\chi}{v_\rho}  c_\beta)$\\
\hline
\end{tabular}
\label{table3cv}
\end{table}

 We have discussed the new contributions that arise in our model concerning the Higgs signal strength so far. In the following we will turn our attention to remaining  scalars, namely $h_2$ and $h^{++}$.

\subsection{ Decay of the heavy scalars, $h_2$ and $h^{++}$}

 Inasmuch as in this work  we are assuming the new scalars of the model, $h_2$ and $h^{++}$ to be heavy, but not heavy enough to decay into the RM331 particles, the total decay width of $h_2$ is determined by the following channels only,
\begin{equation} \label{h2totaldecay}
\Gamma(h_2 \rightarrow all)  =  \Gamma_{(h_2 \rightarrow  \bar{l}l ,\overline{q}q )} + \Gamma_{(h_2 \rightarrow ZZ)}  + \Gamma_{(h_2 \rightarrow WW)} + \Gamma_{(h_2 \rightarrow \gamma \gamma)} + \Gamma_{(h_2 \rightarrow gg)}+\Gamma_{(h_2 \rightarrow h_1h_1)}.
\end{equation}
The decay widths are similar to the $h_1$ case with the couplings  of table~\ref{tableh1} being swapped by those of table~\ref{tableh2}. It is worth pointing  that $h_2$ may decay into a pair of higgs-like particles as well, with a decay with given by,
\begin{equation}
\Gamma(h_2 \rightarrow h_1h_1) \simeq  \frac{1}{8\pi m_{h_2}} (g_{h_2h_1h_1})^2  \left( 1- \frac{4m_{h_1}^2}{m_{h_2}^2} \right)^{\frac{1}{2}}, 
\label{h2->h1h1}
\end{equation}
where, 
$$g_{h_2h_1h_1} =  \frac {\lambda_3 v_{\chi}}{2} \left( c_{\beta}(1 - 3 s_{\beta}) + \frac {v_\rho} {v_\chi} s_{\beta}(1-3 c_{\beta})\right).$$

In regard to the doubly charged scalar,  it is important to notice that it carries two units of lepton number. Thus, as long as it is lighter than the new gauge bosons, it will always decay into only pairs of same equally charged leptons according to Eq.~(\ref{effective}), with the following decay width, 
\begin{equation}
\Gamma(h^{++} \rightarrow l^+ l^+) \sim \dfrac{g^2}{8 \pi} \dfrac{m_{l}^2 m_{h^{++}}}{m^2_W}
 c^2_\alpha
\left( 1 - \dfrac{4m_{l}^2}{m_{h^{++}}^2}\right)^{3/2}.
\label{larguracarregado}
\end{equation}
Observe that there is a lepton mass dependence in the numerator of Eq.~(\ref{larguracarregado}). Hence, one may naively predict that $h^{++}$ decays almost exclusively into  the heaviest lepton pair, namely $\tau^{+}\tau^{+}$. In the next section we perform our numerical analysis based on the results here exposed.
 
\section{Numerical Analysis}
 \label{analysis}

 We first present our numerical results concerning the signal strength of the RM331 model for our Higgs-like scalar, $h_1$.   Next, we compute the branching ratios of the heavy neutral scalar, $h_2$. Lastly, we will calculate the decay width of the doubly charged scalar.

\subsection{Higgs-like scalar}
 Particle physicists are thrilled due the LHC results since last year. The awaited Higgs discovery and the current precise measurements that took place soon after have brought lots of attention mainly due to the arguable diphoton excess. This channel is particularly interesting because many particle physics models predict an enhancement in the diphoton channel. In case this excess is confirmed in the near future that will provide a striking evidence for models where this enhancement can be accounted for. That being said, here we aim to investigate where this model stands concerning this excess and what sort of information we can extract when we plug in the most recent CMS and  ATLAS  measurements regarding the Higgs.

There are two key parameters in our model which play a major role in reproducing the Higgs current measurements, namely $\tan \beta$ , which carries information about the $h_1,h_2$ mixing, and $v_\chi$ that tell us at what energy scale new physics effect beyond RM331 have to come into play. Thus, we will vary these parameters in order to find the best fit choices to LHC  results.

\begin{figure}[!h]
\includegraphics[scale=1.2]{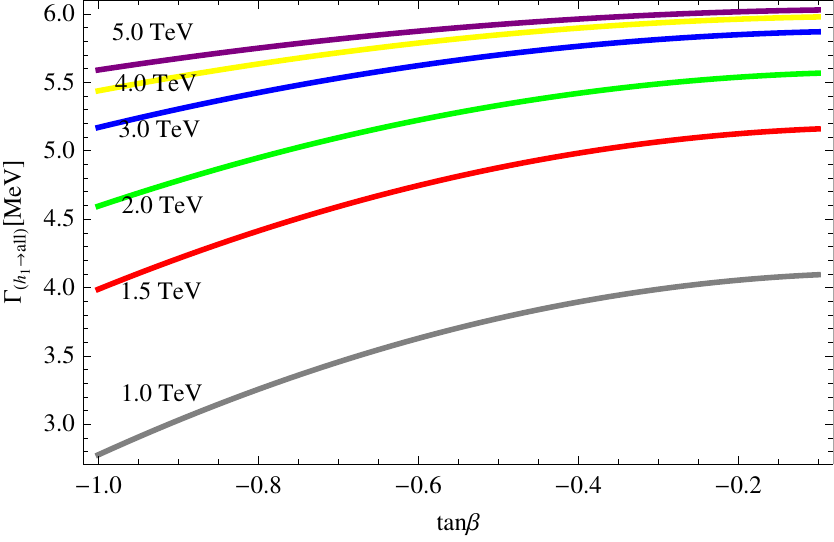}
\caption{Result for the $\Gamma(h_1 \rightarrow all)$ in the  RM331 model as function of the mixing angle $\beta$. }
\label{Gamma-tan}
\end{figure}

We first  start presenting the behavior of the total width of $h_1$ as function of $\tan \beta$ for typical values of $v_\chi$  in FIG. \ref{Gamma-tan}.  The total width decay rate is not well enough measured  to bias some choices of the parameter space. It is understood that the current LHC measurements are pointing to $\Gamma (h\rightarrow all) \sim 6.1$~MeV \cite{vernon}. However, this value suffers from large uncertainties which allow us to conclude that for any choice of parameter in FIG. \ref{Gamma-tan} our model is able to reproduce the total width within the error bars. It is important to emphasize though, that the total decay width  does not reveal the properties of this higgs-like scalar, neither tell us if this scalar is compatible with current measurements for each individual channel. To carry on this investigation we need to face our model with the precise measurements of the signal strength for the higgs boson.

\begin{figure}[!h]
\includegraphics[scale=0.6]{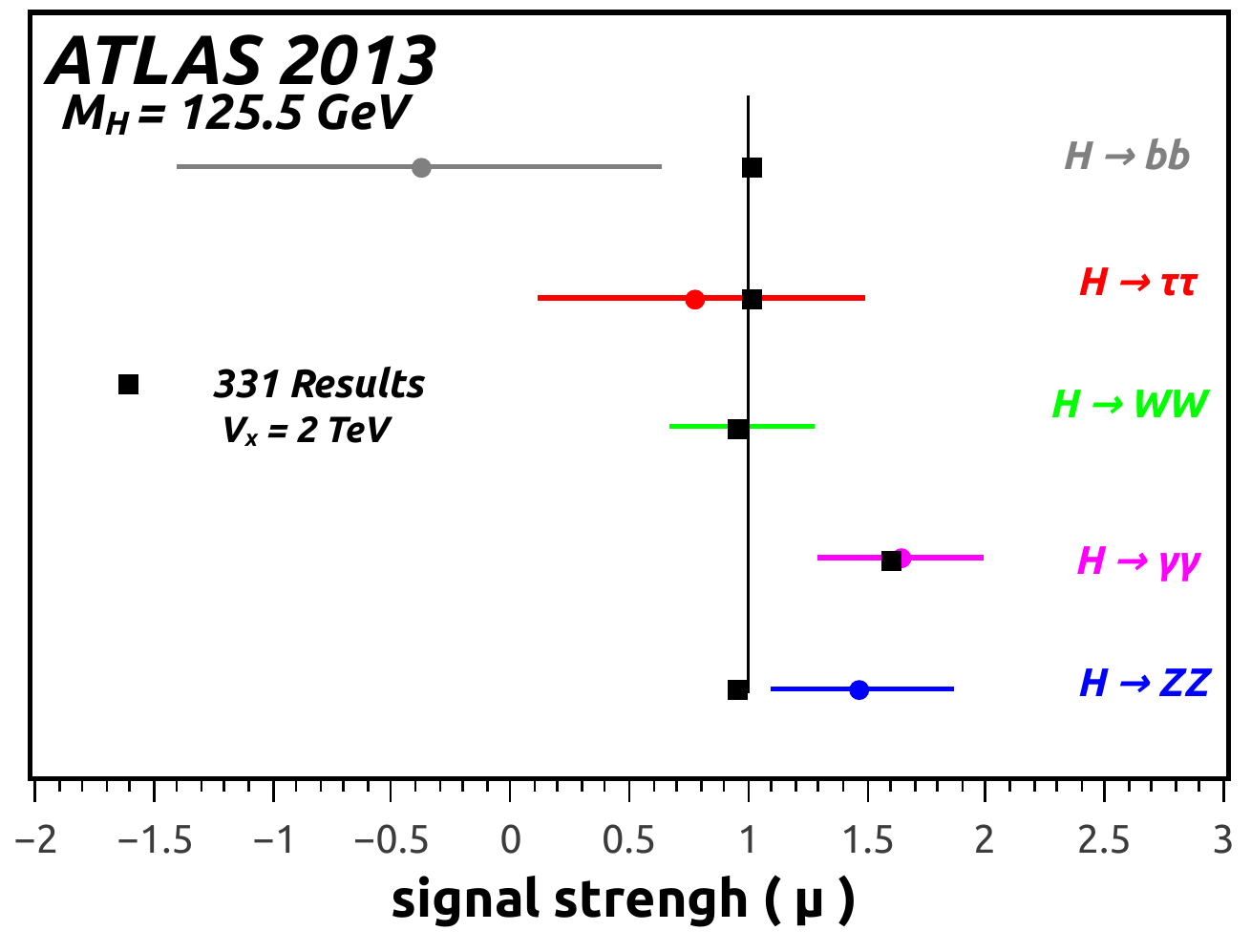}
\caption{Signal strength for all higgs decay channels for   $tan\beta = -0.3227$ and $v_{\chi}=2.0$~TeV . We have included the ATLAS combined data and the RM331 results in the plot.}
\label{ATLAS}
\end{figure}

\begin{figure}[!h]
\includegraphics[scale=0.6]{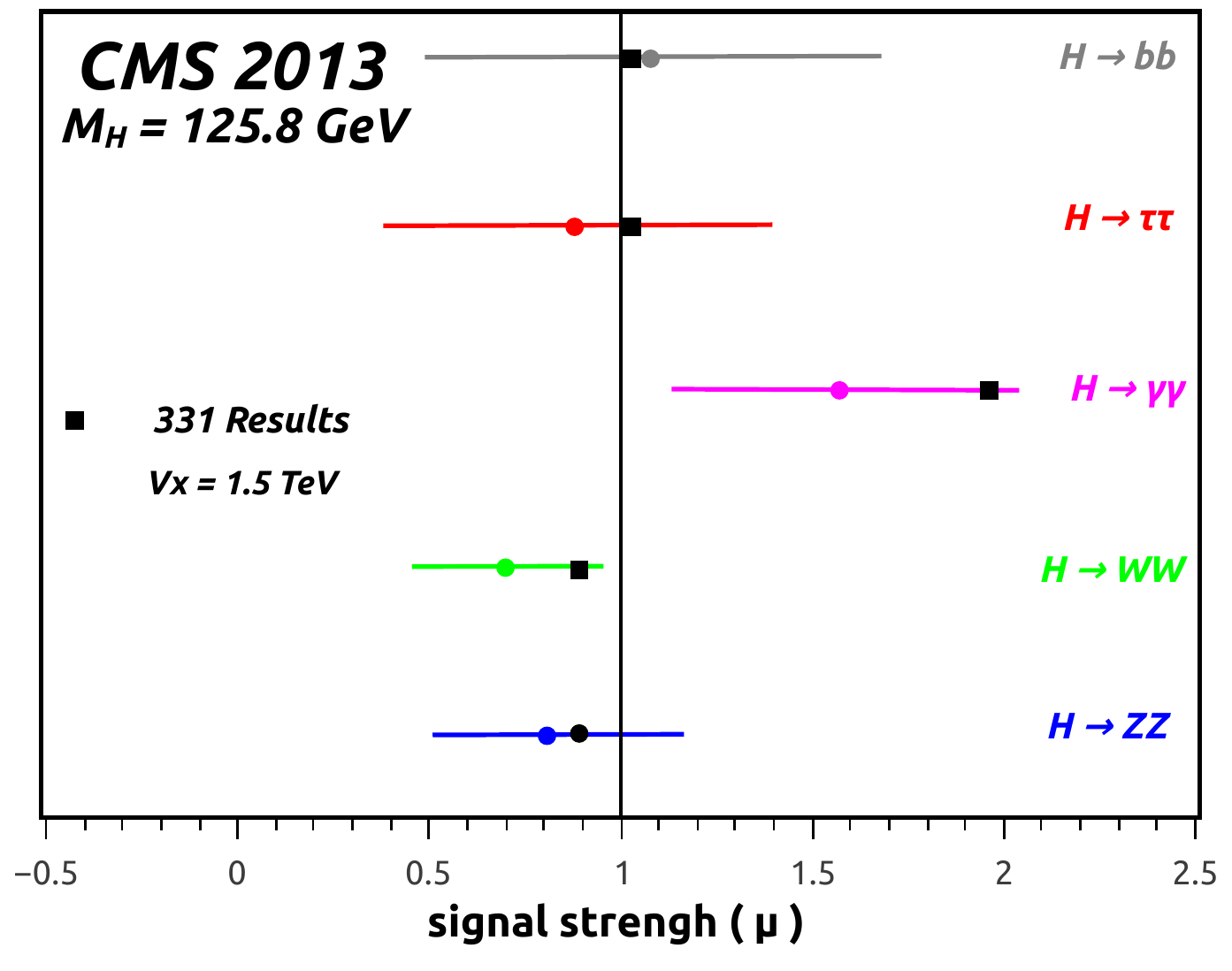}
\caption{Signal strength for all higgs decay channels for $tan\beta = -0.4476$ and $v_{\chi}=1.5$~TeV . In the plot we have included CMS combined data and the RM331 results.}
\label{CMS}
\end{figure}

We  will start with the combined analyses  performed by ATLAS that indicates $\mu^{ATLAS}_{\gamma \gamma} = (1.6 \pm 0.3)$ for $m_h=125.5$~GeV, while the combined CMS one points to $\mu^{CMS}_{\gamma \gamma} = (1.55 \pm 0.45)$ for $m_h=125.8$~GeV, i.e. both combined analyses present a mild excess in the diphoton channel. In view of this, in FIG.~\ref{ATLAS} we show  the signal strength for the combined ATLAS measurements for  $m_{h_1}=125.5$~GeV,  $v_{\chi} \sim 2.0$~TeV and $tan\beta = -0.3227$, while in FIG.~\ref{CMS} we show  the signal strength for the combined CMS measurements for   $m_{h_1}=125.8$~GeV,  $v_{\chi} \simeq 1.5$~TeV and  $tan\beta = -0.4476$.

 Besides those individual results, a recent  global best fit analysis  has been performed including data from Atlas, CMS, D$\emptyset$ and  CDF (Global Analyses)~\cite{ellis}. According to the fit, an excess persists in the diphoton channel and in FIG.~\ref{Global} we show that for different  choices of $v_{\chi}$ and $tan\beta$ our model is able to reproduce this combined report .  

 In summary, our model is able to address either best fit analyses considered here for $tan\beta \simeq -(0.3-0.4)$ and $v_{\chi} \sim 2.0$ roughly, and the higher $v_{\chi}$ the more Higgs-like our scalar $h_1$ is. 

\subsection{Heavy Higgs $h_2$}

In regard to the heavy scalar, its total decay width is found to be composed  of the same channels presented in Eq.~(\ref{totalhiggs}), adjusted  by the respective couplings shown in table~\ref{tableh2}, and by adding the  decay width $\Gamma(h_2 \rightarrow h_1h_1)$.  The  values of the total decay width, $\Gamma_{h_2}$,  for  ($\tan \beta\,,\, v_\chi$)  equal to ($-0.3227\,,\, 2.0$TeV) and ($-0.1556,  4.0$TeV),  are estimated as  $\Gamma_{h_2} \simeq 22.4$~GeV  and $\Gamma_{h_2} \simeq 52.7$~GeV, respectively.

\begin{figure}[!h]
\includegraphics[scale=1.2]{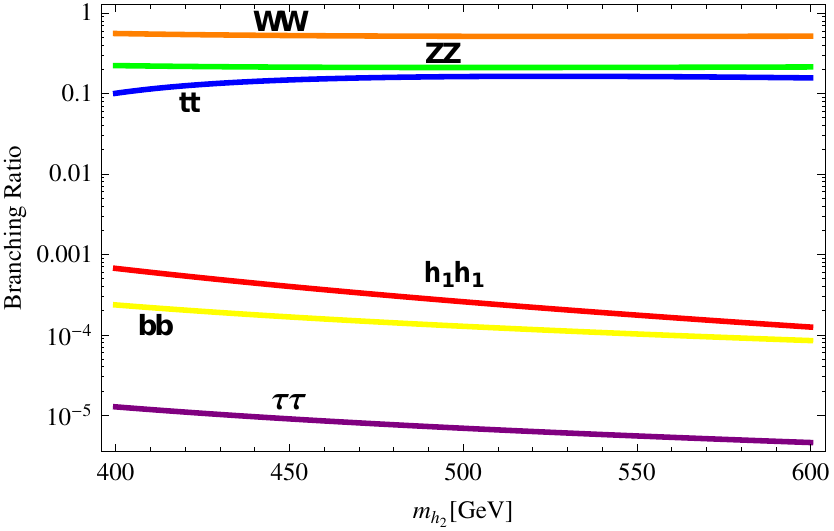}
\caption{Branching Ratio of $h_2$ for all decay channels in the  RM331 model with $v_\chi = 1.0$ TeV .}
\label{BRh2-1TeV}
\end{figure}

In FIG.~\ref{BRh2-1TeV} we show the BR for  $h_2 \rightarrow W^+W^-$, $ZZ$, $h_1h_1$, $\overline{t}t$, $\overline{b}b$ and $\tau^+\tau^-$ channels for $v_\chi = 1.0$~TeV. In particular, one might notice that the WW decay channel is the most important one for a large range of masses and the BR in $h_1 h_1$ is the second most significant decay channel, contributing roughly $20\%$. Interestingly, this BR into Higgs-like particle happens to be much greater than other extended Higgs sector models, such as $U(1)_{B-L}$ SM extension, which predicts a BR into a pair of Higgs of the order of $10^{-8}$~\cite{ubl}. 

\begin{figure}[!h]
\includegraphics[scale=1.2]{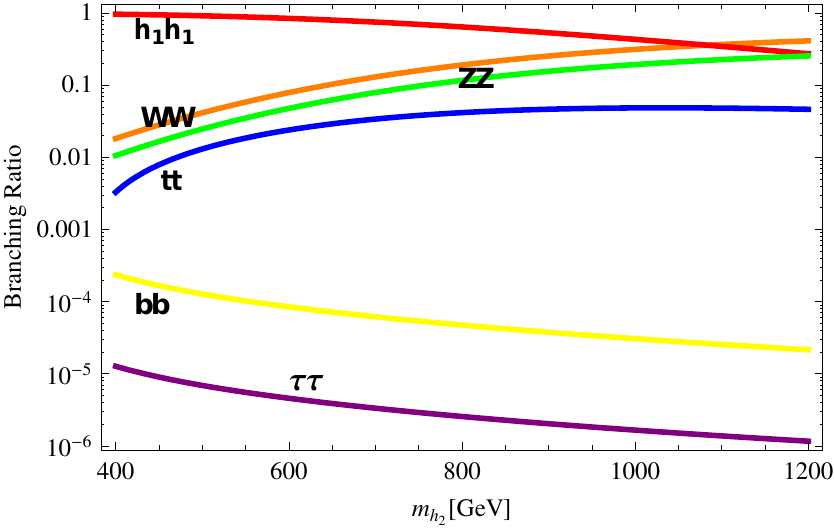}
\caption{Branching Ratio of $h_2$ for all decay channels in the  RM331 model with $v_\chi = 3.0$ TeV.}
\label{BRh2-3TeV}
\end{figure}

\begin{figure}[!]
\includegraphics[scale=1.2]{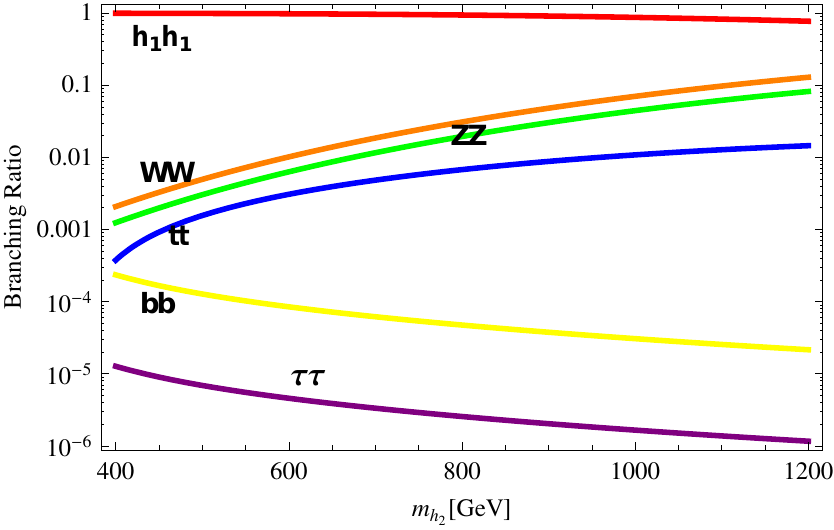}
\caption{Branching Ratio of $h_2$ for all decay channels in the  RM331 model with $v_\chi = 5.0$ TeV.}
\label{BRh2-5TeV}
\end{figure}

The FIGs.~\ref{BRh2-3TeV} and \ref{BRh2-5TeV} contain the results of these BRs for  $v_{\chi} = 3.0$ and $5.0$~TeV, respectively. We observe that $h_2$ decays preferentially into a pair of $h_1$ in these cases. This happens because the trilinear coupling  $h_1h_2h_2$ increases about one order of magnitude when we vary from $v_{\chi} \sim 3.0$~TeV to $v_{\chi} \sim 5$~TeV. 
In short, what we would like to point out is that our heavy scalar $h_2$ decays preferentially into a pair of Higgs with a BR $\geq 90\%$  for $v_\chi \geq 3.0$ TeV, with WW and ZZ being the next most relevant decay channels.

In regard to the  doubly charged scalar, a recent report  about   LHC searches for double charged scalars  that decay almost exclusively into taus, i.e., BR $(h^{++} \rightarrow \tau \tau) \sim 100\%$, put the  following bound in its mass: $m_{ h^{++} } \geq 204$~GeV \cite{doubly}. 
With this in mind, we exhibit in FIG.~\ref{doubly_charged} its decay width into leptons for three different values of $v_\chi = 1.0$, $1.5$ and $5.0$ TeV. Note that it decays preferentially into a pair of tau leptons with $\Gamma(h^{++} \rightarrow \tau^+\tau^+)=(0.9-1.1)$~MeV for $m_{ h^{++} }=(200-300)$~GeV.
With these results we can remark that the RM331 is able to accommodate the observed Higgs-like decay channels in LHC experiments, including some possible deviation from SM concerning the diphoton channel, and also poses specific signatures for the Higgs CP-even partner, $h_2$, very peculiar to this model, with BR into Higgs much higher than one would expect. Also, the doubly charged scalar can show up in the next run of the LHC with a clear signature in pairs of like-charged tau leptons, which can be contrasted with rare leptonic decays of current models.

\section{Conclusions}
\label{conclusions}
\begin{figure}[!h]
\includegraphics[scale=0.6]{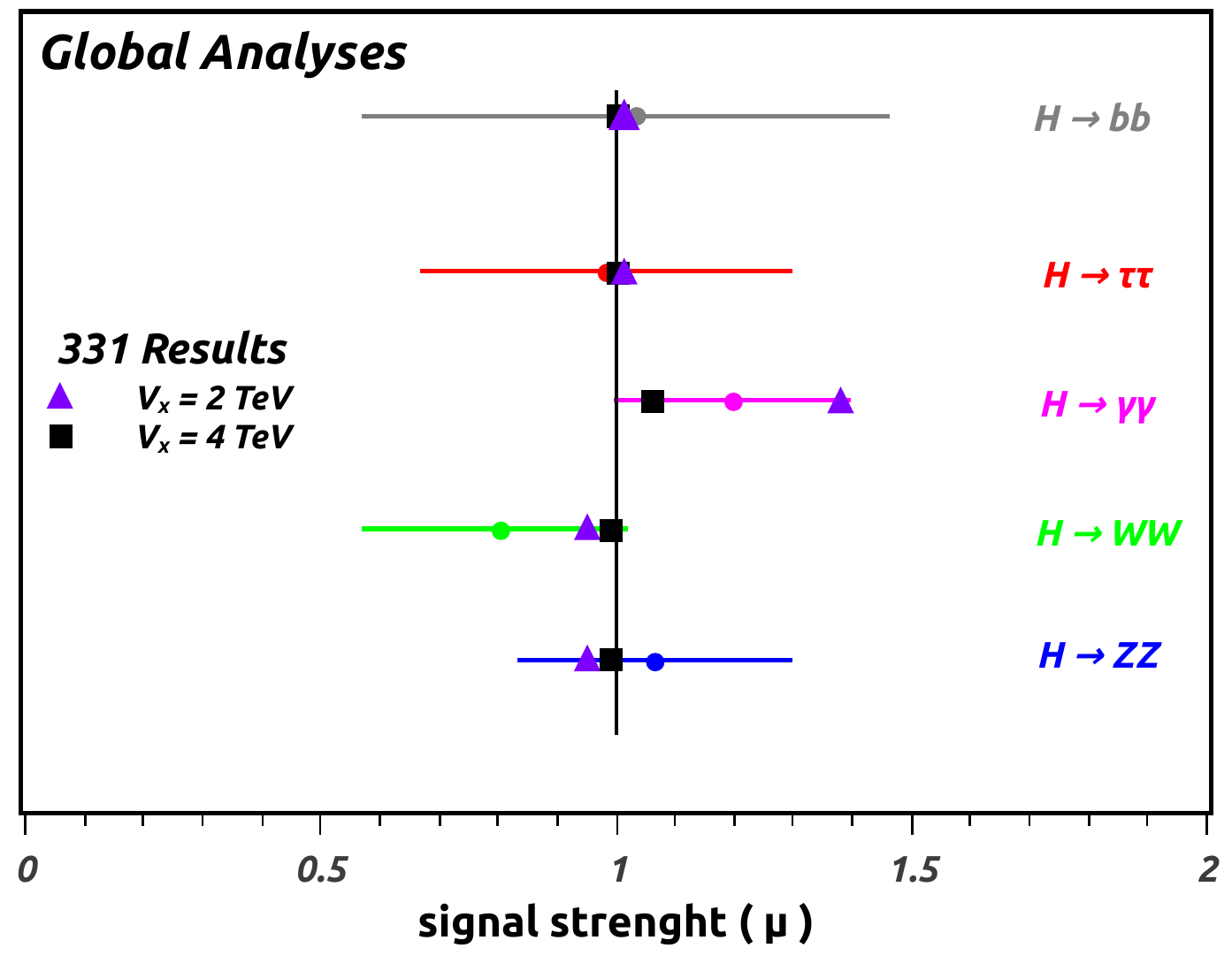}
\caption{Signal strength for all higgs decay  channels in the RM331 model. In the  plot we have included global analyses best fit values and the RM331 results for two different values of the parameters $\tan \beta$ and $v_\chi$: ( -0.3227\,,\, 2.0TeV) and  ( - 0.1556\,,\, 4.0TeV), respectively }
\label{Global}
\end{figure}

\begin{figure}[!h]
\includegraphics[scale=1.2]{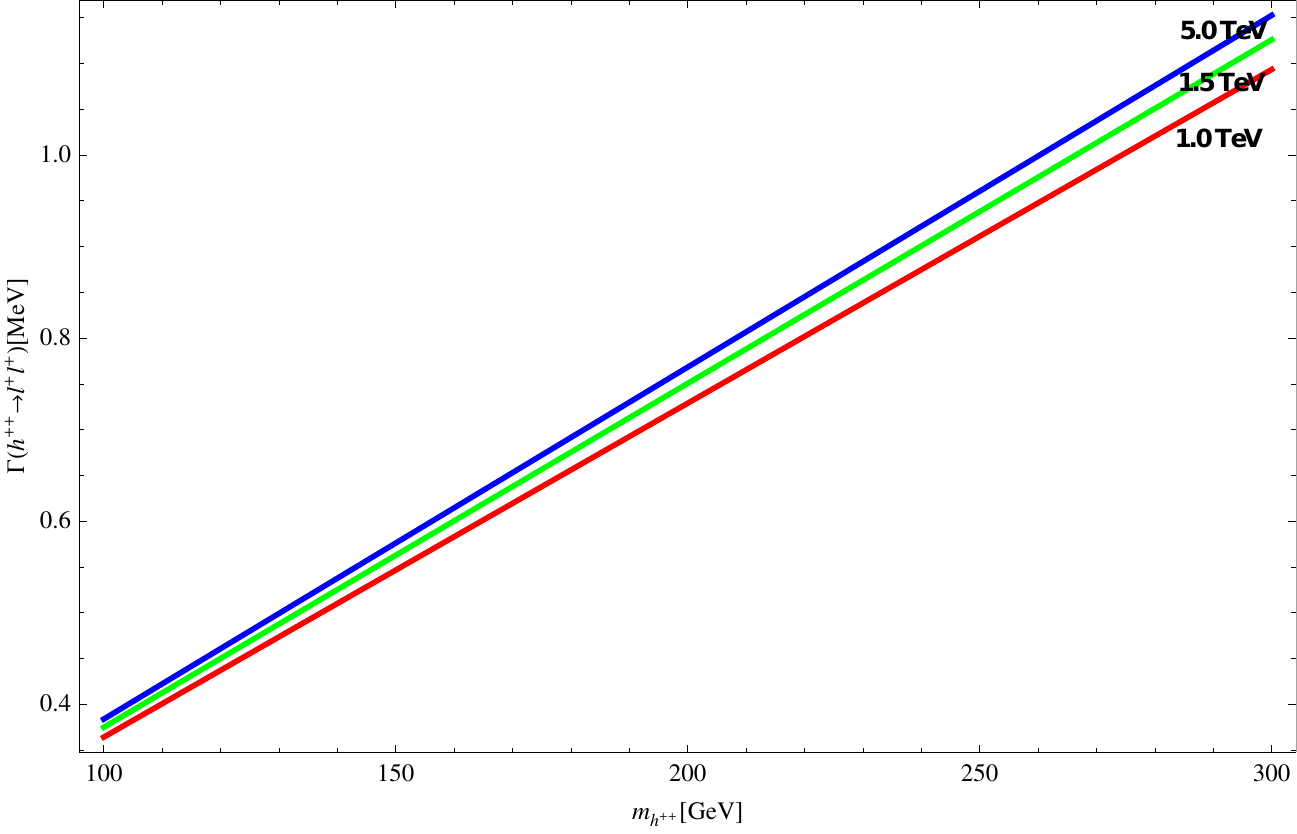}
\caption{The total width decay of the doubly charged scalar into leptons  $h^{++} \rightarrow \ell^+ \ell^+$ for $v_{\chi}$ = 1.0\,,\,1,5 and 5.0 TeV}
\label{doubly_charged}
\end{figure}

The RM331 model is a gauge extension of the SM with a reduced scalar sector composed  by two neutral scalars and one doubly charged one. In this work we developed  some  phenomenology of the scalar  sector of this model. We derived all partial decay widths for the lightest scalar, which is recognized as our Higgs-like particle and  confronted them with the  global analyses which has been performed after gathering the data from ATLAS, CMS, D$\emptyset$ and CDF experiments. We  obtained a good fit to the data as can be seen in FIG. (\ref{Global}). In particular, with $v_{\chi}=2$~TeV we can remarkably account for the $h \rightarrow \gamma \gamma$ excess while still being completely consistent with all the other channels within the error bars. Additionally, we  have computed the branching ratio of the neutral scalar $h_2$  and found that it decays preferentially  into a pair of Higgs-like particles with a branching ratio $\geq 90\%$, a feature not easily obtained in other extensions of SM. We also obtained analytically the total width of the doubly charged scalar, $h^{++}$, predicted in the model and exhibited its behavior  in FIG.~\ref{doubly_charged}. We saw that  $h^{++}$ decays almost exclusively into a pair of tau leptons. This feature may be probed  at the next run of LHC and represents a characteristic signal of this model. In conclusion, we studied the scalar sector of a new version of the 331 model and  showed that at a scale of a few TeV this model is a compelling alternative to the SM once it is able to explain the recent measurements of LHC  regarding the signal strength.

\begin{acknowledgements}
WC is supported by Coordena\c{c}\~ao de Aperfei\c{c}oamento de Pessoal de N\'{\i}vel Superior (CAPES). CASP, FSQ and PSRS are supported by Conselho Nacional de Pesquisa e Desenvolvimento 
Cient\'{\i}fico - CNPq. The authors would like to thank Patricia Telles, Alexandre Alves, Alex Dias and Patrick Draper for useful discussions.
\end{acknowledgements}
\section*{Appendix }
\label{appendix}
After SSB, the mass matrix for the up quark in the basis $u=(u_1,u_2,u_3)$ is,
\begin{eqnarray}
M^{u} &=& \begin{pmatrix} 
\lambda_{11}^{u}\frac{v_\rho v_\chi}{2\Lambda} \ & \lambda_{12}^{u}\frac{v_\rho v_\chi}{2\Lambda}  & \lambda_{13}^{u}\frac{v_\rho v_\chi}{2\Lambda}\\
-\lambda_{21}^{u}\frac{v_\rho}{\sqrt{2}}  & -\lambda_{22}^{u}\frac{v_\rho}{\sqrt{2}}  & -\lambda_{23}^{u}\frac{v_\rho}{\sqrt{2}} \\
-\lambda_{31}^{u}\frac{v_\rho}{\sqrt{2}}  & -\lambda_{32}^{u}\frac{v_\rho}{\sqrt{2}}  & -\lambda_{33}^{u}\frac{v_\rho}{\sqrt{2}} \\ 
\end{pmatrix},
\end{eqnarray}
while the mass matrix for the down quarks in the basis $d=(d_1,d_2,d_3)$ is,
\begin{eqnarray}
M^{d} &=& \begin{pmatrix} 
\lambda_{11}^{d}\frac{v_\rho}{\sqrt{2}} \ & \lambda_{12}^{d}\frac{v_\rho}{\sqrt{2}}  & \lambda_{13}^{d}\frac{v_\rho}{\sqrt{2}}\\
\lambda_{21}^{d}\frac{v_\rho v_\chi}{2\Lambda}  & \lambda_{22}^{d}\frac{v_\rho v_\chi}{2\Lambda}  & \lambda_{23}^{d}\frac{v_\rho v_\chi}{2\Lambda} \\
\lambda_{31}^{d}\frac{v_\rho v_\chi}{2\Lambda}  & \lambda_{32}^{d}\frac{v_\rho v_\chi}{2\Lambda}  & \lambda_{33}^{d}\frac{v_\rho v_\chi}{2\Lambda} \\ 
\end{pmatrix},
\end{eqnarray}

The interactions among the neutral scalars $h_1$  and $h_2$ and the quarks in the flavor basis are,
\begin{eqnarray}
\label{lagrangian}
\mathcal{L} & = & \overline{u}_L\Gamma_1^uu_R h_1 + \overline{u}_L\Gamma_2^uu_R h_2 \nonumber\\
            &   & + \overline{d}_L\Gamma_1^dd_R h_1 + \overline{d}_L\Gamma_2^dd_R h_2 + h.c,
\end{eqnarray} 
where,
\begin{eqnarray}
&&\Gamma_1^u = \dfrac{M^{u}}{v_{\rho}}c_\beta - s_\beta F^u, \,\,\,\,\, \Gamma_1^d = \dfrac{M^{d}}{v_{\rho}}c_\beta - s_\beta F^d, \nonumber \\
&& \Gamma_2^u = \dfrac{M^{u}}{v_{\rho}}s_\beta + c_\beta
F^u, \,\,\,\,\, \Gamma_2^d = \dfrac{M^{d}}{v_{\rho}}s_\beta + c_\beta 
F^d.
\end{eqnarray}
with,
\begin{eqnarray}
F^u =  
\left(\begin{array}{ccc} 
\lambda_{11}^{u}\frac{v_\rho}{2\Lambda} & \lambda_{12}^{u}\frac{v_\rho}{2\Lambda} & \lambda_{13}^{u}\frac{v_\rho}{2\Lambda}\\
0 & 0 & 0 \\
0 & 0 & 0 \\ 
\end{array}\right),\,\,\,\,\mbox{and}\,\,\, F^d =  
\left(\begin{array}{ccc} 
0 & 0 & 0\\
\lambda_{21}^{d}\frac{v_\rho}{2\Lambda} & \lambda_{22}^{d}\frac{v_\rho}{2\Lambda} & \lambda_{23}^{d}\frac{v_\rho}{2\Lambda} \\
\lambda_{31}^{d}\frac{v_\rho}{2\Lambda} & \lambda_{32}^{d}\frac{v_\rho}{2\Lambda} & \lambda_{33}^{d}\frac{v_\rho}{2\Lambda} \\ 
\end{array}\right).
\end{eqnarray}

In the case $\Gamma^{u,d}  \propto M^{u,d}$, both matrices can be diagonalized simultaneously avoiding flavour changing neutral currents (FCNC) processes mediated by $h_1$. Otherwise, $h_1$ may mediate FCNC processes, because the three families are not in the same representation, as a result of triangle anomalies cancellation requirement.  Anyway, without loss of generality, we do not take into account the FCNC processes by considering $M^{u,d}$ and $\Gamma^{u,d}$ diagonal. Under these assumptions the interactions  among quarks and physical scalars are presented in tables~\ref{tableh1} and \ref{tableh2}, for  $q^{\prime}=u,s,b$ and $q=d,c,t$.
\begin{table}[h]
\centering
\caption{Higgs-like ($h_1$) Interactions.}
\begin{tabular}{|c|c|}
\hline 
Interactions & Couplings\\
\hline
$\overline{l}lh_1$ & $\dfrac{m_l}{v_{\rho}}\left(c_\beta - \frac{v_\rho}{v_\chi}s_\beta\right)$ \\
$\overline{q}qh_1$   & $\dfrac{m_q}{v_{\rho}}c_{\beta}$ \\ 
$\overline{q'}q'h_1$ & $\dfrac{m_q'}{v_{\rho}}\left(c_\beta - \frac{v_\rho}{v_\chi}s_\beta\right)$ \\
$W^+W^-h_1$          &  $\frac{1}{2} g^2v_{\rho}c_{\beta}$\\
$ZZh_1$ 				   & $\frac{1}{4} g^2v_{\rho}\sec^{2}_{\theta_W}c_{\beta}$\\ 
$V^{+}V^{-}h_1$      & $-\frac{1}{2} g^2 v_\chi s_\beta$\\
$U^{++} U^{--}h_1$   & $\frac{1}{2} g^2 v_\rho (c_\beta  - \frac{v_\chi}{v_\rho}  s_\beta)$\\
$h^{++} h^{--}h_1$   & $\lambda_4 v_\rho (c_\beta  - \frac{v_\chi}{v_\rho}  s_\beta)$\\
\hline 
\end{tabular}
\label{tableh1}
\end{table}


\begin{table}[h]
\centering
\caption{Heavy Scalar ($h_2$) Interactions.}
\begin{tabular}{|c|c|}
\hline 
Interactions & Couplings\\
\hline
$\overline{l}lh_2$ & $ \dfrac{m_l}{v_{\rho}} \left(s_\beta + \frac{v_\rho}{v_\chi}c_\beta\right)$ \\
$\overline{q}qh_2$   & $\dfrac{m_q}{v_{\rho}}s_{\beta}$ \\ 
$\overline{q'}q'h_2$ & $\dfrac{m_q'}{v_{\rho}}\left(s_\beta +\frac{v_\rho}{v_\chi}c_\beta\right)$  \\
$W^+W^-h_2$          &  $\frac{1}{2} g^2v_{\rho}s_{\beta}$\\
$ZZh_2$ 				   & $\frac{1}{4} g^2v_{\rho}\sec^{2}_{\theta_W}s_{\beta}$\\ 
$V^{+}V^{-}h_2$      & $\frac{1}{2} g^2 v_\chi c_\beta$ \\
$U^{++} U^{--}h_2$   & $\frac{1}{2} g^2 v_\rho (s_\beta  + \frac{v_\chi}{v_\rho}  c_\beta)$\\
$h^{++} h^{--}h_2$   & $\lambda_4 v_\rho (s_\beta  + \frac{v_\chi}{v_\rho}  c_\beta)$\\
\hline
\end{tabular}
\label{tableh2}
\end{table}

\newpage





\newpage

\begin{thebibliography}{}
%
\bibitem{atlas}
 G. Aad \textit{et al}. (Atlas Collaboration), Phys. Lett.  {\bf B716},  (2012) 1, [arXiv:1207.7214].
\bibitem{cms} 
S. Chatrchyan \textit{et al}. (CMS Collaboration), Phys. Lett. {\bf B716},  (2012) 30, [arXiv:1207.7235].
\bibitem{explaining} 
A. Alves \textit{et al}, Phys. Rev. {\bf D84} 115004 (2011), [arXiv:1109.0238]; J. Chang, K.  Cheung, P-Y. Tseng, T-C. Yuan,  JHEP 1212 ,(2012) 058, [arXiv:1206.5853]; A. Alves \textit{et al}. Eur. Phys. J. C {\bf 73}, 2288 (2013), [arXiv:1207.3699]; J. R. Espinosa, C. Grojean, M. Muhlleitner, M. Trott,  JHEP 1212, (2012) 045, [arXiv:1207.1717]; J. Cao, Z. Heng, J. M. Yang, J. Zhu,  JHEP 1210, (2012) 079,[arXiv:1207.3698]; A.  Delgado, G. Nardini, M. Quiros, Phys. Rev.  D{\bf 86}, (2012) 115010,[arXiv:1207.6596]; A. Urbano, Phys. Rev.  D{\bf 87},  (2013) 053003,[arXiv:1208.5782]; A. Alves, Phys. Rev. D{\bf 86},  (2012) 113010, [arXiv:1209.1037]; Z.  Chacko, R.  Franceschini, R. K. Mishra, JHEP 1304, (2013) 015, [arXiv:1209.3259];  E.  Bertuzzo, P. A.N. Machado, R. Zukanovich Funchal, JHEP 1302, (2013) 086, [arXiv:1209.6359]; M. Drees,  Phys. Rev. D{\bf 86}, (2012) 115018, [arXiv:1210.6507] ; Lei Wang, Xiao-Fang Han, Phys. Rev. D 87 (2013) 015015, [arXiv:1209.0376].
\bibitem{RencontresdeMoriond} Moriond Proceeding contributions, http://moriond.in2p3.fr/QCD/2013/MorQCD13Prog.html
\bibitem{331-1} F. Pisano and V. Pleitez, Phys. Rev. D{\bf 46},(1992) 410,[hep-ph/9206242].
\bibitem{331-2} P. H. Frampton, Phys. Rev. Lett. {\bf 69},(1992) 2889.
\bibitem{ecq}
C. A. de S. Pires, O. P. Ravinez, Phys. Rev. D{\bf 58},  (1998) 035008,[hep-ph/9803409]; C. A. de S. Pires,  Phys. Rev. D{\bf 60},  (1999) 075013,[hep-ph/9902406].

\bibitem{neutrinos1} A. G. Dias, C. A. de S. Pires, P. S. Rodrigues da Silva, Phys. Lett. B{\bf 628}, (2005) 85,[hep-ph/0508186]; F. Queiroz, C. A. de S. Pires, P. S. Rodrigues da Silva, Phys. Rev. D{\bf 82} ,(2010) 065018,[arXiv:1003.1270].
\bibitem{families} D. Cogollo, H. Diniz, C.A. de S.Pires, P.S. Rodrigues da Silva, Mod.Phys.Lett. A {\bf23} (2009) 3405,[arXiv:0709.2913].
\bibitem{DM331}  C. A. de S. Pires, F. S. Queiroz, P. S. Rodrigues da Silva, Phys. Rev. D{\bf 82}, (2010) 105014,[arXiv:1002.4601]; J. K. Mizukoshi, C. A. de S. Pires, F .S. Queiroz, P. S. Rodrigues da Silva, Phys.Rev. D{\bf 83}, (2011) 065024,[arXiv:1010.4097]; J. D. Ruiz-Alvarez, C. A. de S.Pires, Farinaldo S. Queiroz, D. Restrepo, P. S. Rodrigues da Silva, Phys. Rev. D{\bf 86}, (2012) 075011,[arXiv:1206.5779].
\bibitem{higgs331} 
See the first and the last papers in Ref. \cite{explaining}.
\bibitem{novelsourses} J. A. Rodriguez, M. Sher, Phys.Rev. D{\bf 70}, (2004) 117702,[hep-ph/0407248];  C. Promberger, S. Schatt, F.  Schwab,  Phys. Rev. D{\bf 75}, (2007) 115007,[hep-ph/0702169]; R. H. Benavides, Y. Giraldo, William A. Ponce, Phys.Rev. D{\bf 80}, (2009) 113009,[arXiv:0911.3568]; D. Cogollo \textit{et al}.  Eur. Phys. J. C{\bf 72} (2012) 2029,[arXiv:1201.1268]; A.C.B. Machado, J.C. Montero, V. Pleitez, [arXiv:1305.1921].
 
\bibitem{exotic} E. Ramirez Barreto, Y.A. Coutinho, J. Sa Borges, Phys.Lett. B{\bf 689}, (2010) 36-41,[ arXiv:1004.3269]; E. Ramirez Barreto, Y. A. Coutinho, J. Sa Borges, Phys.Rev. D{\bf 83}, (2011) 075001,[arXiv:1103.1267]; A. Alves, E. Ramirez Barreto, A. G. Dias, Phys. Rev. D{\bf 84},  (2011) 075013,[arXiv:1105.4849].
\bibitem{dong} Y.  Giraldo, W.  A. Ponce, Eur. Phys. J. C{\bf 71}, (2011) 1693,[arXiv:1107.3260]; A. Doff and A. A. Natale, Int.J.Mod.Phys. A{\bf 27},  (2012) 1250156,[arXiv:1210.3390];P. V. Dong, T. D. Tham, H. T. Hung, [arxiv:1305.0369].
\bibitem{paulodias} J. G.  Ferreira, Jr, P. R. D.  Pinheiro, C. A. de  S.  Pires, P. S. Rodrigues da Silva, Phys. Rev. D{\bf84}, (2011) 095019,[arXiv:1109.0031].
\bibitem{pheno}
 For some phenomenology of this model, see: V. T. N. Huyen, T. T.  Lam, H. N. Long, V. Q. Phong, [arXiv:1210.5833]. For the supersymmetric version of this model, see: D. T. Huong, L.T. Hue, M. C. Rodriguez, H. N. Long, Nucl. Phys.B {\bf 870}, (2013) 293-322,[arXiv:1210.6776];
\bibitem{landau}A. G. Dias, R. Martinez, V. Pleitez,  Eur. Phys. J. {\bf C 39}(2005), 101, [hep-ph/0407141]. See also,  A. G. Dias, V. Pleitez, Phys. Rev. {\bf D80} (2009), 056007,[arXiv:0908.2472].
\bibitem{rizzo} Thomas Rizzo, Phys. Rev. {\bf D22}, (1980) 722.
\bibitem{GeneralL} J. Ellis, M. K. Gaillard and D. Nanopoulos, Nucl. Phys. B{\bf 106},(1976) 292; A. I. Vainshtein, M. B. Voloshin, V. I. Zakharov and M. A. Shifman, Sov. J. Nucl. Phys. {\bf 30}, (1979) 711.
\bibitem{vernon} V. Barger, M. Ishida, W-Y Keung, Phys. Rev. Lett. {\bf 108}.  (2012) 261801,[arXiv:1203.3456].
\bibitem{ellis} T. Aaltonen {\it et al.} (CDF Collaboration, D0 Collaboration), Phys. Rev. Lett. {\bf 109}, (2012) 071804,[arXiv:1207.6436]; John Ellis and Tevong You, [arxiv:1303.3879].
\bibitem{ubl} W. Emam and S. Khalil, Eur. Phys. J.  C{\bf 55} (2007), 625,[arXiv:0704.1395].
\bibitem{doubly} CMS Collaboration Eur. Phys. J. C{\bf 72} (2012), 2189,[arXiv:1207.2666].
\end{thebibliography}
\end{document}